\documentclass{sigchi-ext}
\usepackage[T1]{fontenc}
\usepackage{textcomp}
\usepackage[scaled=.92]{helvet} %
\usepackage{graphicx} %
\usepackage{balance}  %
\usepackage{booktabs} %
\usepackage{ccicons}  %
\usepackage{ragged2e} %
\newcommand{\para}[1]{\noindent {\bf #1}}

\def\plaintitle{Harnessing Explanations to Bridge AI and Humans} 

\def\emptyauthor{}
\def\plainkeywords{explanations, interpretable machine learning, human-centered explanations}

\title{Harnessing Explanations to Bridge AI and Humans}

\numberofauthors{3}
\author{%
  \alignauthor{%
    \textbf{Vivian Lai}\\
    \affaddr{University of Colorado Boulder} \\
    \affaddr{Boulder, CO, USA} \\
    \email{vivian.lai@colorado.edu} }\vfil \alignauthor{%
    \textbf{Samuel Carton}\\
    \affaddr{University of Colorado Boulder}\\
    \affaddr{Boulder, CO, USA} \\
    \email{samuel.carton@gmail.com} } \vfil \alignauthor{%
    \textbf{Chenhao Tan}\\
    \affaddr{University of Colorado Boulder} \\
    \affaddr{Boulder, CO, USA} \\
    \email{chenhao@chenhaot.com} }}

\definecolor{linkColor}{RGB}{6,125,233}
\hypersetup{%
  pdftitle={\plaintitle},
  pdfauthor={\emptyauthor},
  pdfkeywords={\plainkeywords},
  bookmarksnumbered,
  pdfstartview={FitH},
  colorlinks,
  citecolor=black,
  filecolor=black,
  linkcolor=black,
  urlcolor=linkColor,
  breaklinks=true,
}

\begin{document}

\CopyrightYear{2020}
\setcopyright{rightsretained}
\conferenceinfo{CHI'20,}{April  25--30, 2020, Honolulu, HI, USA}
\isbn{978-1-4503-6819-3/20/04}
\doi{https://doi.org/10.1145/3334480.XXXXXXX}
\copyrightinfo{\acmcopyright}

\maketitle

\RaggedRight{} 

\begin{abstract}
Machine learning models are increasingly integrated into societally critical applications such as recidivism prediction and medical diagnosis, thanks to their superior predictive power. 
In these applications, however, full automation is often not desired due to ethical and legal concerns. 
The research community has thus ventured into developing 
interpretable methods that 
explain 
machine predictions.
While these explanations are meant to assist humans in understanding machine predictions and thereby allowing humans to make better decisions, this 
hypothesis is not 
supported in many recent studies.
To improve human decision-making with AI assistance, we propose future directions 
for closing the gap between the efficacy of explanations and improvement in human performance.

\end{abstract}

\section{Why Do We Need Explanations?}
Recent trends in machine learning have 
led to
models that are increasingly powerful, complex, opaque, and ubiquitous. 
Model performance has begun to meet or exceed expert human performance in numerous areas such as 
recidivism prediction \cite{kleinberg2017human} and medical diagnosis \cite{ardila_end--end_2019}. 
Concomitantly, AI models have begun to play a larger and larger role in aspects of life such as government, business, and science, leading to ever-higher consequences for model mistakes.  

Unfortunately, while average model performance has approached human levels, models still lag behind humans in key ways. AI models 
tend to absorb bias from their training data, %
are vulnerable to adversarial inputs, %
and have difficulty generalizing beyond the specific distribution of that training data
\cite{guidotti_survey_2018}.
A common suggestion to mitigate these issues is to view models as \textbf{augmenting} rather than replacing human effort. In the ideal scenario, a human and a model could work together as a hybrid system whose performance would exceed that of either agent operating alone. 
AI explanations have been proposed as a way to achieve this cooperation. The %
hypothesis
is that if 
a human can scrutinize the logic behind a model prediction, they can recognize when that prediction is unfair, nonsensical, or otherwise unreliable \cite{guidotti_survey_2018}. 
\section{The Current State of Explanations}

In order to achieve a balance between AI accuracy and human intuition, the research community has proposed a number of techniques for explaining the predictions of AI models. 
A common approach is feature attribution, which attempts to assign each feature (word, 
pixel, etc.) a score indicating its importance in the model's prediction. 
Such methods range from retroactive perturbation-based analysis like the popular LIME \cite{ribeiro2016should} to built-in attention mechanisms such as that proposed by Lei et al. (2016) \cite{lei_rationalizing_2016}. 

However, the community has struggled to demonstrate an improvement in human decision quality as a result of these kinds of explanations. 
Typical experimental design involves human subjects making decisions in the presence of model predictions and evaluating whether explanations improve their accuracy in doing so. Some experiments in this vein have included 
predicting apartment prices \cite{poursabzi-sangdeh_manipulating_2018}, 
detecting deceptive online reviews \cite{lai+tan:19},
assessing social media toxicity \cite{carton_attention-based_2020}, 
performing various artificial tasks \cite{lage2019evaluation}, 
and recidivism prediction \cite{green2019principles}.
We are not aware of any such experiment that has reported a significant improvement in accuracy that cannot be explained by increased subject trust in a model whose accuracy is 
higher than the human baseline (such as Lai and Tan 2019 \cite{lai+tan:19}).

Why have explanations failed to improve human performance? 
While this is a difficult question to answer, existing results provide a few clues. First, 
Lai et al. (2020) point out two distinct types of AI learning problem: \textit{emulating} human skill vs. \textit{discovering} new knowledge  \cite{lai+liu+tan:20}.
They speculate that in the latter case, humans may not have strong enough task intuitions to make effective use of simple explanations, leading to a need for additional training \cite{lai+liu+tan:20}.
Even in emulation tasks, models may incidentally learn patterns that simply do not correspond well with human intuition, as was observed  by Feng et al. (2018) in the case of LSTM models for sentiment analysis \cite{feng2018pathologies}. 
Explanations may be better suited for catching certain type of model errors over others: Carton et al. (2020) observe that they reduce false positives while increasing false negatives, surmising that subjects find it easier to overturn phrases incorrectly identified as toxic than to discover truly toxic phrases missed by the model \cite{carton_attention-based_2020}.

Overall, these results suggest a fundamental misalignment between AI explanations and human mental models, a situation that Bansal et al. (2019) discuss as a general hurdle in human-AI collaboration \cite{bansal2019beyond}. 
As a solution, we suggest two basic directions for future work: 1) 
augmenting human mental models to cope with model explanations; and 2) 
adjusting model explanations and behavior to match human mental models.

\section{Direction I: Augmenting Human Mental Models}

\para{Model-driven tutorials.} 
Humans seem to not have strong intuition in making effective use of explanations in tasks %
that \emph{discover} new knowledge \cite{lai+liu+tan:20}.
To improve human mental models, we propose model-driven tutorials that elucidate counter-intuitive and inconspicuous patterns embedded in models learned from the dataset.
Model-driven tutorials are one possible way to align human mental models and AI, and we call for more study on how to effectively train humans to work with AI explanations.

\para{Interactive explanations.} 
The goal of interactive explanations is to allow humans to understand the model better through trial-and-error scenarios.
As compared to static explanations that only reveal what is important to the model, interactive explanations allow humans to interact with models and explanations, e.g., by editing input and examining the differences in a model's prediction.
Instead of simply presenting important patterns in the model,
it is 
useful for humans to identify patterns through active learning.

\para{Evaluating generalization.} 
It is important to point out that a typical setup in prior work employs a random split to obtain training and testing data, which is a standard assumption in supervised machine learning. 
While humans can ideally improve generalization in this case, humans might be more likely to correct generalization errors in machine learning models when the testing distribution differs from training. 
In that case, understanding the embedded patterns, especially spotting spurious ones, can help humans generalize these data-driven insights and \emph{reduce model biases}. 
A significant challenge lies in how we can properly evaluate such generalization, relating to a core issue in machine learning.

\section{Direction II: Towards Human-Centered Explanations}

\para{Understanding human explanations.} 
Existing techniques tend to optimize explanations for numeric qualities like sparsity or some notion of fidelity to the model. Ultimately, however, 
we need to recognize that explanations serve as a communicative device to humans.
Key to this idea is more effort to understand the rationales behind human decisions, the qualities of those rationales associated with correct and incorrect decisions, and the effect of model-human rationale alignment on model-human agreement.
Studies such as Kaushik et al. 2019 \cite{kaushik_learning_2019} which collect human rationales/explanations are a good start, but we call for a {\bf behavioral} and {\bf design} perspective on such data rather than its use merely as additional training signal. 

\para{Experimenting with alternative explanation types.}
Feature attribution may simply be inadequate for affording meaningful human oversight of model predictions, especially in discovery-type tasks where they don't have strong existing intuitions.
Example-based explanations
and 
natural language explanations
may succeed where feature-based explanations fail. Therefore, we call for more human subject experimentation involving alternative explanation styles.

\para{Explanations as model criticism.} Another focus area we suggest is to break away from treating explanations as a diagnostic signal for the reliability of a static model. Perhaps instead we should treat them as a means for critiquing the underlying logic of model decisions that are known to be incorrect. While the idea of ``learning from explanations'' has a long history \cite{zaidan_using_2007}, we are not aware of work that employs this idea in a dynamic way, in response to known model errors, and which incorporates existing model explanations.

\section{Conclusion}

AI explanations have generated great excitement as a way to 
provide added value
in high-stakes decision-making.
However, they have been failing in recent studies to live up to their promise. We suggest new research directions to address this expectation gap, based on the idea of aligning AI and human mental models to enable the type of critical human scrutiny that is likely to lead to real improvements.   

\balance{} 

\bibliographystyle{SIGCHI-Reference-Format}
\bibliography{ref}


\begin{thebibliography}{00}


\ifx \showCODEN    \undefined \def \showCODEN     #1{\unskip}     \fi
\ifx \showDOI      \undefined \def \showDOI       #1{{\tt DOI:}\penalty0{#1}\ }
  \fi
\ifx \showISBNx    \undefined \def \showISBNx     #1{\unskip}     \fi
\ifx \showISBNxiii \undefined \def \showISBNxiii  #1{\unskip}     \fi
\ifx \showISSN     \undefined \def \showISSN      #1{\unskip}     \fi
\ifx \showLCCN     \undefined \def \showLCCN      #1{\unskip}     \fi
\ifx \shownote     \undefined \def \shownote      #1{#1}          \fi
\ifx \showarticletitle \undefined \def \showarticletitle #1{#1}   \fi
\ifx \showURL      \undefined \def \showURL       #1{#1}          \fi

\bibitem{ardila_end--end_2019}
{Diego Ardila}, {Atilla~P. Kiraly}, {Sujeeth Bharadwaj}, {Bokyung Choi},
  {Joshua~J. Reicher}, {Lily Peng}, {Daniel Tse}, {Mozziyar Etemadi}, {Wenxing
  Ye}, {Greg Corrado}, {David~P. Naidich}, {and} {Shravya Shetty}. 2019.
\newblock \showarticletitle{End-to-end lung cancer screening with
  three-dimensional deep learning on low-dose chest computed tomography}.
\newblock {\em Nature Medicine\/} (2019).
\newblock


\bibitem{bansal2019beyond}
{Gagan Bansal}, {Besmira Nushi}, {Ece Kamar}, {Walter~S Lasecki}, {Daniel~S
  Weld}, {and} {Eric Horvitz}. 2019.
\newblock \showarticletitle{Beyond Accuracy: The Role of Mental Models in
  Human-AI Team Performance}. In {\em AAAI}.
\newblock


\bibitem{carton_attention-based_2020}
{Samuel Carton}, {Qiaozhu Mei}, {and} {Paul Resnick}. 2020.
\newblock \showarticletitle{Attention-{Based} {Explanations} {Don}’t {Help}
  {Humans} {Detect} {Misclassifications} of {Online} {Toxicity}}. In {\em
  ICWSM}.
\newblock


\bibitem{feng2018pathologies}
{Shi Feng}, {Eric Wallace}, {II Grissom}, {Mohit Iyyer}, {Pedro Rodriguez},
  {and} {Jordan Boyd-Graber}. 2018.
\newblock \showarticletitle{Pathologies of neural models make interpretations
  difficult}. In {\em EMNLP}.
\newblock


\bibitem{green2019principles}
{Ben Green} {and} {Yiling Chen}. 2019.
\newblock \showarticletitle{The principles and limits of algorithm-in-the-loop
  decision making}. In {\em CSCW}.
\newblock


\bibitem{guidotti_survey_2018}
{Riccardo Guidotti}, {Anna Monreale}, {Franco Turini}, {and} {Dino Pedreschi}.
  2018.
\newblock \showarticletitle{A Survey Of Methods For Explaining Black Box
  Models}.
\newblock {\em {ACM Computing Surveys}\/} (2018).
\newblock


\bibitem{kaushik_learning_2019}
{Divyansh Kaushik}, {Eduard Hovy}, {and} {Zachary~C. Lipton}. 2020.
\newblock \showarticletitle{Learning the Difference that Makes a Difference
  with Counterfactually-Augmented Data}. In {\em ICLR}.
\newblock


\bibitem{kleinberg2017human}
{Jon Kleinberg}, {Himabindu Lakkaraju}, {Jure Leskovec}, {Jens Ludwig}, {and}
  {Sendhil Mullainathan}. 2017.
\newblock \showarticletitle{Human decisions and machine predictions}.
\newblock {\em The quarterly journal of economics\/} (2017).
\newblock


\bibitem{lage2019evaluation}
{Isaac Lage}, {Emily Chen}, {Jeffrey He}, {Menaka Narayanan}, {Been Kim}, {Sam
  Gershman}, {and} {Finale Doshi-Velez}. 2019.
\newblock \showarticletitle{An evaluation of the human-interpretability of
  explanation}.
\newblock {\em arXiv:1902.00006\/} (2019).
\newblock


\bibitem{lai+liu+tan:20}
{Vivian Lai}, {Han Liu}, {and} {Chenhao Tan}. 2020.
\newblock \showarticletitle{"Why is 'Chicago' deceptive?" Towards Building
  Model-Driven Tutorials for Humans}. In {\em CHI}.
\newblock


\bibitem{lai+tan:19}
{Vivian Lai} {and} {Chenhao Tan}. 2019.
\newblock \showarticletitle{On Human Predictions with Explanations and
  Predictions of Machine Learning Models: A Case Study on Deception Detection}.
  In {\em FAT*}.
\newblock


\bibitem{lei_rationalizing_2016}
{Tao Lei}, {Regina Barzilay}, {and} {Tommi Jaakkola}. 2016.
\newblock \showarticletitle{Rationalizing {Neural} {Predictions}}. In {\em
  EMNLP}.
\newblock


\bibitem{poursabzi-sangdeh_manipulating_2018}
{Forough Poursabzi-Sangdeh}, {Daniel~G. Goldstein}, {Jake~M. Hofman},
  {Jennifer~Wortman Vaughan}, {and} {Hanna Wallach}. 2018.
\newblock \showarticletitle{Manipulating and Measuring Model Interpretability}.
\newblock {\em arXiv:1802.07810 [cs]\/} (2018).
\newblock


\bibitem{ribeiro2016should}
{Marco~Tulio Ribeiro}, {Sameer Singh}, {and} {Carlos Guestrin}. 2016.
\newblock \showarticletitle{Why should i trust you?: Explaining the predictions
  of any classifier}. In {\em KDD}.
\newblock


\bibitem{zaidan_using_2007}
{Omar Zaidan}, {Jason Eisner}, {and} {Christine Piatko}. 2007.
\newblock \showarticletitle{Using "Annotator Rationales" to Improve Machine
  Learning for Text Categorization}. In {\em NAACL}.
\newblock


\end{thebibliography}

\end{document}